\documentstyle[psfig]{caosp}
\begin{document}
\def\Teff{$T_{\rm eff}$}
\def\logg{$\log{g}$}
\def\logZ{[M/H]}
\def\Vsini{$v\cdot\sin{i}$}
\def\Vmicro{$v_{\rm micro}$}
\def\kms{\,km\,s$^{-1}$}
\pubyear{1997}
\volume{23}
\firstpage{1}
\htitle{Chemical composition and fundamental parameters of roAp stars}
\hauthor{M.J. Gelbmann}
\title{Chemical composition and fundamental parameters of roAp stars}
\author{M.J. Gelbmann}
\institute{Institute for Astronomy, T\"urkenschanzstr. 17, A-1180 Vienna, Austria}
\date{December 28, 1997}
\maketitle
\begin{abstract}
Element abundances of three roAp stars, HD\,166473, HD\,203932, and HD\,217522, 
were determined using Kurucz
model atmospheres with metal abundances scaled to solar ones and the results 
were compared with data from the literature concerning three further roAp
stars, normal B and A stars and two $\lambda$ Bootis stars. Up to 38 elements
could be identified and therefore, this work represents the most complete 
chemical investigation hitherto published, which can be summarized as follows: 

\begin{description}
\item{$\bullet$} all investigated roAp stars have a similar abundance pattern, 
\item{$\bullet$} the overabundances of rare earth and other heavy elements are
comparable to cool non-pulsating Ap-stars, 
\item{$\bullet$} iron belongs to the most deficient and cobalt to the most
enhanced elements in the group of the iron peak elements, and 
\item{$\bullet$} the light elements carbon, nitrogen, and oxygen are less
abundant than in atmospheres with abundances scaled to the Sun. 
\end{description}

\noindent Beside an unexpected possible relation between effective temperature
and metallicity of roAp stars, no outstanding differences from non-pulsating Ap
stars could be detected. This statement, however, suffers from the lack of
comparably detailed investigations of the latter.

\end{abstract}


\section{Introduction}

Rapidly oscillating Ap (roAp) stars are a subgroup of the CP2 stars, which 
oscillate with non-radial, high overtone, low order acoustic $p$-modes with 
the axis of oscillation aligned with the axis of the magnetic field 
(Kurtz 1982). A still open question is concerned with the excitation 
mechanism for roAp stars and other physical parameters which distinguish 
them from non-pulsating CP stars. 

\section{Observations}

Six EMMI echelle spectra were obtained by G.\,Mathys with the NTT 
at ESO in June 1992. Two different cross-disperser grisms 
covered the entire spectral range from 4200 to 8100\,\AA. The spectral 
resolution is about {$\cal R$}\,=\,24000, the typical
signal-to-noise ratio per pixel is about 200 in the continuum.

All reductions of the observations to continuous spectra and the normalization
to the continuum were done within the echelle package of IRAF. 
All calculations to determine the basic stellar parameters and the abundance
patterns are based on the
Abundance Analysis Procedure AAP (Gelbmann 1995).

\section{Fine Analyses}

An abundance analysis was carried out for the three roAp 
stars HD\,166473, HD\,203932 (Gelbmann et al. 1997), and HD\,217522. 
Abundance analyses of three further roAp stars have been 
carried out by the Vienna working group with the same spectrum synthesis 
tool: $\alpha$\,Cir (Kupka et al. 1996), $\gamma$\,Equ (Ryabchikova et al. 
1997), and HD\,24712 (Ryabchikova et al. 1997a).

To identify the abundance peculiarities of the group of roAp stars, their 
abundance spectra were compared to four normal B stars (Adelman 1986) and 15 
early A type stars (Hill 1995).
In addition, abundance analyses of two $\lambda$\,Bootis stars with effective
temperatures comparable to the roAp stars have been carried out within the 
Vienna working group with the same spectrum synthesis toolkit (Heiter 1996).

\section{Results}

The effective temperature range of all six roAp stars spectroscopically
investigated so far varies from \Teff\,=\,6750\,K to 8000\,K, corresponding to
spectral types from F2 to A4. The surface gravities range from \logg\,=\,4.2 to
4.4 which is consistent with $\log g$ values for A4 to F2 main sequence
stars in a theoretical HR-diagram (\logg\,=\,4.3, Gray 1992). Although the
accuracy of the determination of \logg\ is rather poor, it is evident that roAp
stars are positioned on or close to the ZAMS, they do not seem to be
considerably evolved. The close proximity of the roAp stars to the ZAMS
indicates that chemical peculiarities have developed on relatively short
time scales. Table\,\ref{parameters} summarizes the fundamental atmospheric 
parameters of all six roAp stars discussed within this study.

\begin{table}
\begin{center}
\begin{small}
\caption{The atmospheric parameters of all roAp stars analyzed so far;
	\logZ\ is the total metallicity of all iron peak elements.
	The stars are sorted in order of increasing temperature.}
\label{parameters}
\begin{tabular}{llccccc} \hline
Star       & Name          & \Teff & \logg & \logZ  & \Vmicro & \Vsini \\
           &               & [K]   &       &        & [\kms]  & [\kms]\\ \hline
HD\,217522 & BP\,Gru       & 6750  & 4.3   &$-$1.36 & 1.0     & 12.0   \\
HD\,24712  & DO\,Eri       & 7250  & 4.3   &$-$0.34 & 1.0     & 7.0    \\
HD\,203932 & BI\,Mic       & 7450  & 4.3   & +0.03  & $<$0.6  & 12.5   \\
HD\,201601 & $\gamma$\,Equ & 7750  & 4.2   & +0.06  & 1.0     & $<$4.5 \\
HD\,128898 & $\alpha$\,Cir & 7900  & 4.2   &  0.00  & 1.5     & 12.5   \\
HD\,166473 & V694\,CrA     & 8000  & 4.4   & +0.67  & 1.0     & 18.0  \\ \hline
\end{tabular}
\end{small}
\end{center}
\end{table}

Although the abundance of atmosphere models for the syntheses are solar scaled 
within a wide range of different scaling factors, all roAp stars 
spectroscopically investigated as yet show similar abundance patterns: 

\begin{table}
\begin{small}
\caption{Abundances of elements normalized to the total number of atoms
	for all six roAp stars and for the Sun (Anders \& Grevesse 1989).} 
\psfig{figure=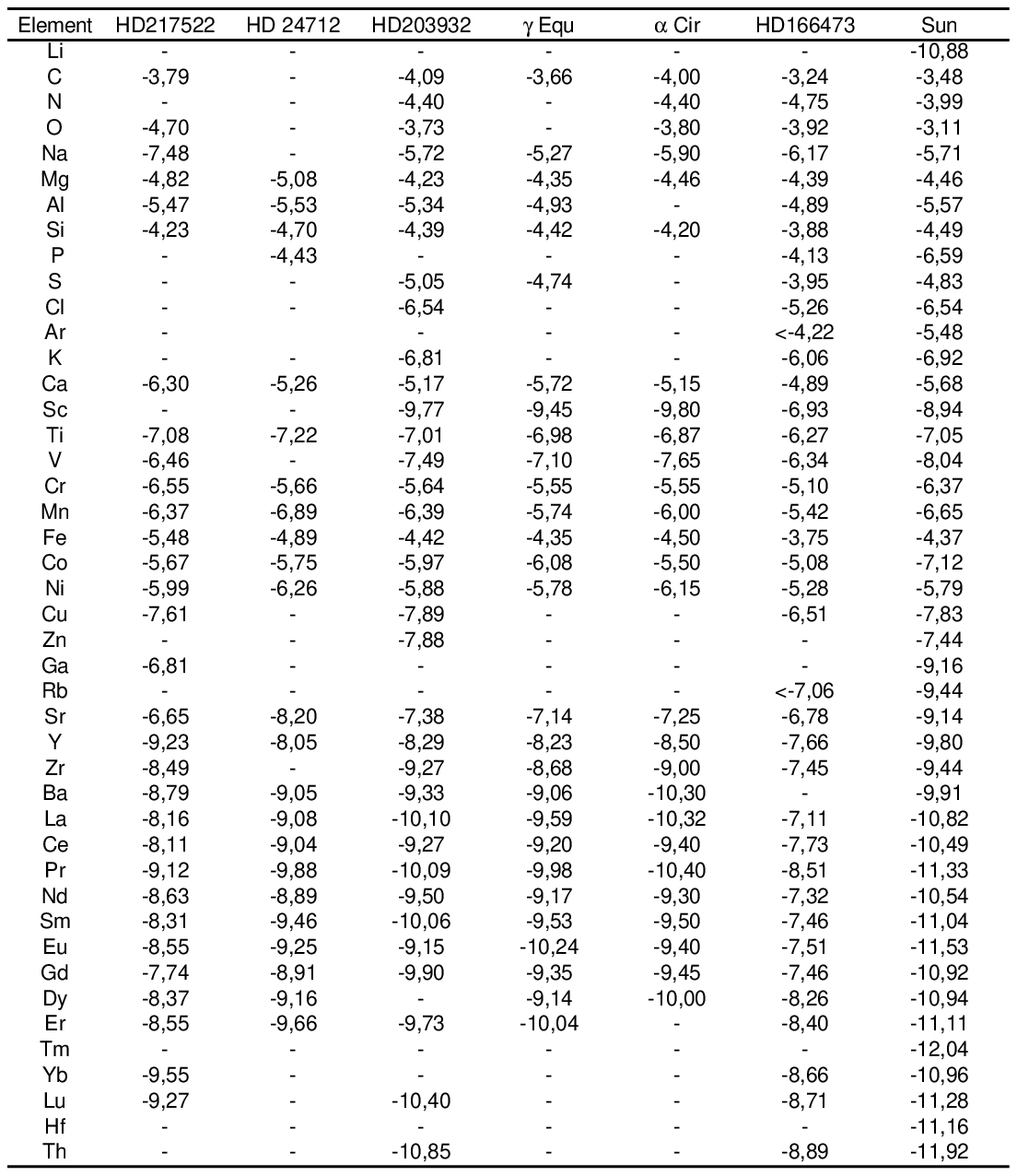,height=13cm}
\label{abund}
\end{small}
\end{table}

The overabundance of rare earth elements and some other heavy elements are
comparable to other cool Ap stars. The elements C, N, and O are less abundant
than the scaled solar abundances, their overall abundances relative to the Sun
vary from [CNO]\,=\,-0.8 to -0.2 with the mean at -0.5.
Most iron peak elements are more abundant than iron, which is the main
contributor to opacity in the atmosphere. For all iron peak elements the total
abundances vary from [I.P.]\,=\,-0.8 to +0.7 with the mean at +0.2, while the
Co abundances vary from [Co]\,=\,+1.0 to +2.0 with the mean at +1.6.
As the abundances of the iron peak elements increase, the abundances of the
heavy elements increase as well. Among normal A-type stars this tendency of
the heavy elements to be overabundant has been noted before by Lemke (1990).
The total abundances of all rare earth elements vary from [REE]\,=\,+1.3 to
+3.4 with the mean at +2.7! Table\,\ref{abund} lists all abundances of the six 
roAp stars in order of increasing temperatures. 

\section{Conclusions}


This study shows that roAp stars have similar abundances for up to 38 
elements identified in the sample. These abundances are being used to compute 
stellar atmosphere models based on individual opacities. Within the 
significantly smaller number of analyzed elements in the literature, a 
comparison to non-roAp stars do not reveal large abundance differences. 
The derived fundamental stellar parameters and abundances allow to locate 
the stars in the HR-diagram and provide important boundary values for 
pulsation models. Stellar structure and evolutionary parameters can be 
derived from such models. 

\begin{figure}
\centerline{
\psfig{figure=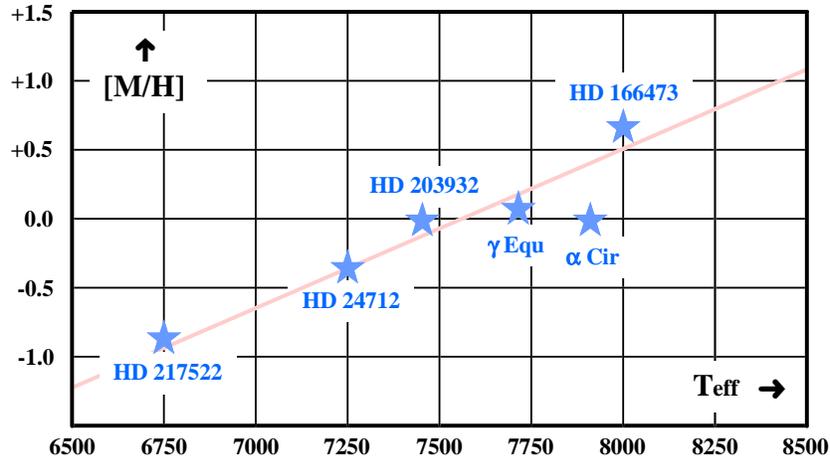,height=6.5cm}}
\caption{Relation between effective temperature and total metallicity
	of all iron peak elements.}
\label{relation}
\end{figure}

A comparison of the iron peak abundances in the six roAp stars
shows an unexpected relation between effective 
temperature and metallicity. 
Fig.\,\ref{relation} shows this tendency for the total metallicity 
of all iron peak elements \logZ\ versus the spectroscopically derived effective
temperature \Teff. However, since all stellar 
parameters -- including also effective temperature -- are determined mainly 
with iron lines, it is not clear whether this tendency is astrophysically 
significant or an artifact due to the applied analyzing algorithm. The 
derived stellar parameters and abundances are resulting from optimization 
routines. Inadequate model atmospheres, limited 
signal-to-noise ratio of the spectra, errors in atomic parameters and 
problems in defining the continuum can result in fairly large errors. 
Therefore, the correlation between metallicity and effective temperature 
has to be further investigated. 

Beside this possible relation, no outstanding differences between pulsating 
and non-pulsating Ap stars could be detected. 



\acknowledgements
I am grateful to Tanya Ryabchikova of the Russian Academy of Sciences 
for contributing with many helpful discussions. I thank Gautier Mathys for 
providing me with observations.

This research was done within the working group {\sl AMS - Asteroseismology
along the Main Sequence}, supported by the Fonds zur F\"orderung der
wissenschaftlichen Forschung (project {\sl P\,8776-AST}). The calculations
were performed with a DEC AXP-3000/600, funded through the Digital Equipment
Corporation project STARPULS.


\end{document}